\documentclass{sig-alternate-10pt}
\usepackage{amsmath}
\usepackage{graphicx}
\usepackage{booktabs}
\usepackage{amsmath}
\usepackage{mdwlist}
\usepackage{printlen}
\usepackage[final]{microtype}
\usepackage[T1]{fontenc}
\usepackage{lmodern}

\begin{document}
\title{Adding Geographical Embedding to AS Topology Generation}
\numberofauthors{2} 
%
\author{
%
%
\alignauthor
Arne Schwabe\\
\affaddr{University of Paderborn}\\
\affaddr{Warburger Stra{\ss}e 100}\\
\affaddr{33098 Paderborn, Germany}\\
\email{arne.schwabe@uni-paderborn.de}
\alignauthor
Holger Karl\\
\affaddr{University of Paderborn}\\
\affaddr{Warburger Stra{\ss}e 100}\\
\affaddr{33098 Paderborn, Germany}\\
\email{holger.karl@uni-paderborn.de}
}
\date{\today}


\maketitle
\hyphenation{au-tono-mous}
\overfullrule=1mm

\begin{abstract}
To study large-scale effects on the Internet various models have been introduced to generate Internet-like autonomous system (AS) topologies. The models for large-scale AS topologies have been focused on replicating structural graph properties. One of the most promising model is the Positive Feedback Model model (PFP). These models however lack the ability to generate routing path and realistic latency. 

We present a model to enrich the AS peering graph with peering points. Our new model allows to calculate path for the connections between end hosts and to infer the latency from these paths. We introduce a new notion for the generation of AS topologies: the compactness of an AS. 

We introduce an algorithm based on the PFP algorithm which generates instances for our model. Verifying the generated model instances shows that the resulting latencies as well as the geographic properties match measured data sets. 
\end{abstract}

\section{Introduction}
\label{sec:introduction}

When designing and evaluating new protocols or applications, it is
import to make realistic assumptions about underlying
infrastructure and its influence on the protocol or application
metrics. For example, when developing latency-sensitive applications
for the Internet, latency characteristics of the Internet should be
well understood. The growing interest in such applications -- for
example, streaming media services, VoIP, interactive cloud-based
services or P2P -- makes practical models for Internet latency highly
desirable, be it for prediction or simulation purposes. 

Latency in the Internet is influenced by routing decisions as well as
by queuing and signal propagation delays,                                                             the latter depends on the
actual physical distances between routers. Existing Internet models
either concentrate on generating a model for the Internet topology,
from which routing models can be derived, but ignore latencies; or
they take an empiric approach and model only latency, without any
recognition of the underlying routing substrate
(Section~\ref{sec:related-work} gives an overview). We maintain that
it is desirable to express both topological aspects and latency in a
single model, allowing to calculate latency of \emph{paths}, not only
links, directly from this one model. Such an integrated model has
advantages over separate models; for example, changing the routing in
the model should affect the path and ultimately the latency. 

In this paper, we propose such an integrated model.  Section \ref{sec:model}
describes the model itself. We present our novel idea of the AS compactness of an AS for the graph
generation and details of our the algorithm to generate model
instances in Section~\ref{sec:as-topology-creation}. 
We evaluate these instances (in
Section~\ref{sec:eval-gener-graph}) by showing that crucial
properties the latencies obtained from our generated topologies match
those observed in widely used datas sets; as properties, we consider
the cumulative distribution function and the triangle inequality
violations.

\section{Related work}
\label{sec:related-work}

We distinguish here between network topology models and empiric
latency models.

\subsection{Empiric latency models}
\label{sec:empir-latency-models}

PeerfactSim \cite{pearfactsim} simulates network latencies as part of its P2P simulation. The simulation of these network latencies is done by using the GNP \cite{Ng2002a} network coordinates. The behavior of the simulated network latencies is subject to the design of the network coordinates. 
Coordinate systems in general aim to predict latencies using as little communication overhead as possible and intentionally trade inaccuracy against efficient computation and low resource usage. Due to this tradeoff some phenomena, however, are not reproduced by a coordinate system. For example, many coordinate systems (GNP among them) do not reproduce triangle violations \cite{Donnet2010}. 
This inaccuracy is justified by proponents of such approaches as being too minor to matter and impossible to recreate with limited knowledge anyway. Consequently, these properties of real networks are not reflected in simulations based on such models. 

Instead of focusing on predicting latency from a limited view at each node the model  by Kaune et al \cite{modelidelay} uses a global view to generate latency that can be used for simulation. Their approach is aimed a producing a latency function that has similar latency characteristics to the data set. Although the goal of the model is very different, the resulting model used to generate the latency is similar to the coordinate system approach. 
The hosts of the CAIDA data set are embedded into a low-dimensional euclidean vector space minimising the quadratic latency error between the metric of the vector space and the CAIDA da set.
To simulate the latency between two hosts the hosts are mapped to the vector space. The metric distance between the hosts is the latency. To account for jitter and other effects not expressed the vector a random component is added to the latency.

\subsection{Topology models}
\label{sec:topology-models}

The previous models did not infer the latency from a network model but used an unrelated model to reproduce the latency, often based on a vector space. A very high abstraction level of the Internet is the peering graph (or the AS topology) which consists of all AS and edges been peering ASes.

Considerable research has been invested in researching the AS topology and in finding algorithms to generate peering graphs. \cite{netgensurvey}. Many properties of the AS peering graph are unique and  are not found in smaller networks or when looking at small subset of the peering graph. 
 These unique properties include the small-world/scale-free property, the power law node degree distribution, disassortative mixing (links between nodes of different types of AS are preferred) \cite{largescaleinet} and the rich club \cite{richclub} property. From the Internet topology generation algorithms, the PFP \cite{pfp} algorithm currently reproduces most of these unique proprieties. The PFP algorithm is a good choice for replicating these unique properties of the peering graph. 

Routing on a peering graph by PFP is possible and will result in an AS path. An AS path is a list of the ASes which are involved in routing a packet from one end host to another.
Calculating latencies from an AS path is not possible since the AS path lacks the information necessary to calculate packet propagation times. A three-hop AS path could represent three ASes in America or three ASes involving different continents. 
To calculate a latency from the AS path more (e.g. geographic) information is needed about the ASes and the points where packets are exchanged between ASes.

The PFP only generates connections on the AS peering level. The GT-ITM algorithm \cite{gtitm} models a finer-detailed relationship between ASes and generates multiple links between different ASes. The GT-ITM algorithm focuses on generating good internal and external AS interconnections on the router level for a small number of ASes. Having this focus on the router level, GT-ITM assigns random locations to the routers. Consequently, the latencies resulting from this topology are not realistic. Since GT-ITM focuses on a small number of ASes it does not reproduce the unique properties of the AS peering graph which are important when modelling a large number of ASes relationship. 

Combining an empiric with the PFP model (or another network topology model) to build a model which can generate latency and models the AS topology model seems like an obvious solution, but unfortunately is not straightforward. The empiric models are missing the network structure information (like AS membership of hosts) to map them to the AS peering graph. Mapping a host individually in empiric model and the network would produce both AS information and latency but has the disadvantage that the latency is not influenced by the network topology and AS path and latency have no relationship or connection to each other. Any effect which is based on the interaction between the two model is not reproducible.

\section{Model}
\label{sec:model}
In this section we describe our model. Our model is build as two layered graphs $G$ and $H$. The graph $G$ is AS peering graph. Using the top-level graph $G$ we model the physical AS interconnection graph $H$. The nodes in $H$ describe the interconnections of  AS border routers.

\begin{figure*}
{\centering
\includegraphics*[width=0.9\textwidth]{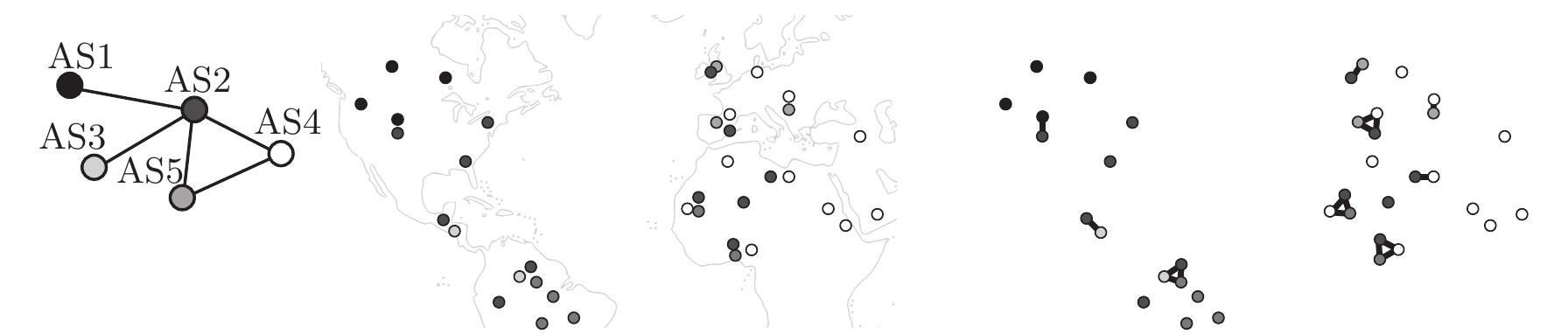}

}
\caption{Relationship of the AS inter-connection graph $G$ (left) and the AS border router graph $H$. The middle figure shows the mapping of AS in  $G$ to locations $l_i$ and the right figure shows the resulting edges between locations (edges between locations of the same AS are not shown)}
\label{figgundh}
\end{figure*}

\subsection{AS level and border router graph}
\label{sec:two-graphs}

The top-level graph AS  $G$ of our model is the AS relationship graph.  This graph is defined as $G=(V,E)$ where $v \in V$ are ASes and $e \in E$ are the edges between peering (connected) ASes. 

The graph $G$ describes the peerings between ASes. To introduce geographic locations to the ASes we define a second graph $H$. We define $C$ as geographic points with latitude and longitude: $C = [-180,180] \times [-90,90]$. For two elements $c_1,c_2 \in C$ we define $|c_1 - c_2|$ as the geographic distance between $c_1$ and $c_2$ (great-circle distance, the shortest distance between two points on the surface of a sphere).

Each AS $v \in V$ is mapped by the function $L$ to a \emph{finite} number of geographic coordinates $l_{v_i} \in C$:  
\begin{equation}\label{lgleichung} L: V \mapsto 2^C, L(v) \rightarrow \{l_{v_1} \ldots l_{v_{n_v}}\} \end{equation}

\noindent where $n_v$ is the number of locations of the AS $v$. For ease of notation we assume that no two ASes are mapped to the same location but arbitrarily close. This simplification allows us to use build the inverse function $L^{-1}: C \mapsto V$, mapping a location to an AS. 

Using $G$ and $L$ induce the graph $H$. $H$ describes the connections between the border routers of ASes. The vertices in $H$ are the border routers of the ASes. 

Let $v_1$ the AS of $l_1$ and $v_2$ the AS of $l_2$: $v_1 = L^{-1}(l_1)$, $v_2= L^{-1}(l_2)$. Two vertices $l_1$ and $l_2$ in the graph $H$ are connected if

\begin{enumerate}
\item\label{con1} Both belong to the same AS: $v_1 = v_2$
\item\label{con2} Both belong to connected ASes (edge $(v_1,v_2)$ $\in E_G$) 
 and are in close proximity $|l_1 - l_2| < L_{\max}$
\suspend{enumerate}
$H$ should represent the same AS peerings as $G$. If a peering between ASes exists in $G$  but not in $H$ an edge in $H$ is added between the two closest locations of the two ASes.

Figure~\ref{figgundh} shows the relationship  of $G$ and
$H$. 
\subsection{Adding latency to the graph}
\label{sec:adding-latency-graph}

The graphs $G$ and $H$ describe our geographically embedded model. So far, it does not specify latency between end devices.  We define $X$ as the set of end devices and the function $\hat d: X \times X \mapsto \mathbb{R}^+$ as \emph{modeled} latency between these devices.

The routing of a network dictates the path of the packets which in turn affects the latency for the packets. The basic routing principle of the Internet and for our function $\hat d$ is:  Each AS tries to minimize its own cost by keeping the packet inside its own network as briefly as possible. Hence, an AS uses greedy routing to forward the packet as quickly as possible to the next AS that has a shorter AS path (hop) distance to the destination. The routing is often referred to as ``hot potato'' routing. When AS optimizes its own path length the overall path length suffers. Our model enables us to reproduce this usually unwanted, yet still observable behavior. Figure~\ref{greedyrouting} shows an example of the effect.

\begin{figure}
\includegraphics{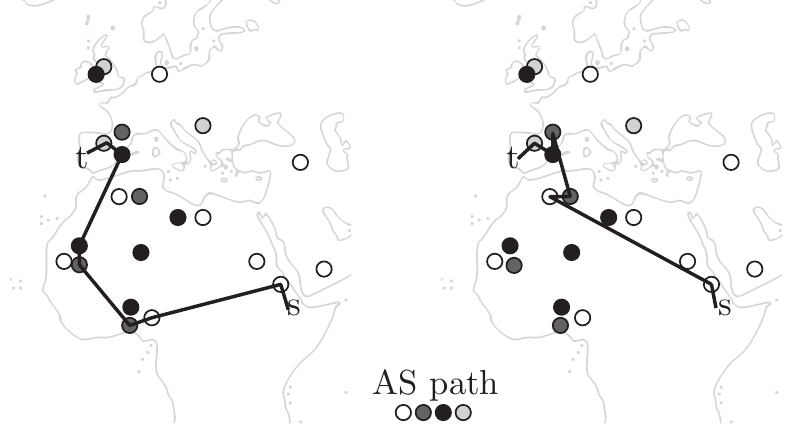}
\caption{Example showing that hot potato from $s$ to $t$ does not necessarily yield the shorted path (left). The path on the right shows routing using the same AS order but a shorter overall path. }\label{greedyrouting}
\end{figure}

We define the function $\hat d$ by means of an algorithm that reflects AS hot potato routing:\begin{enumerate}
\item \label{item:pick-candidate-locations} Let 
$P:X \mapsto 2^C$ be the function which maps end devices to the set of AS locations to which it could be attached; they have to be in close proximity to $x$. 
\[P(x)= \left.\left\{ l_{i} \in V_H \right.\huge\mid\ | h(x) -  l_{i}|  < H_{\max}\right\} \]
\item \label{item:pick-location} For each device, pick one of these attachement options; represent this choice by the function $h: X \mapsto C$ mapping a device $x \in X$ to the location of its point of attachment. 
For any $x$, pick $h(x) \in P(x)$ uniformly at random. 

\item  \label{item:pick-AS-path} To go from a device $x_1$ to a device $x_2$, first define the sequence of ASes that is traversed. The devices define, via their locations $h(x_1)$ and $h(x_2)$,  the ASes to which they are attached, namely $L^{-1}(h(x_i))$; we choose the shortest path in $G$ between them. 

Formally: Construct the shortest path $l_G=(v_1,v_2,$ $\ldots,$ $v_n)$ from $v_1 = L^{-1}(l_{x_1})$ to $v_n=L^{-1}(l_{x_2})$ in $G$. This gives an AS path.
\item \label{item:pick-H-path} Construct a greedy routing path $L_H$ in $H$: Starting with $l_{p_{1,1}}=h({x_1})$ add the node $l_{p_{1,2}} \in L(v_1)$ which has an edge to any node in $L(v_2)$ and has the shortest distance to $l_{p_{1,1}}$. Add a node in $L(v_{2})$ which has an edge to $l_{p_{1,2}}$. Repeat the steps until $l_{p_{n,1}} \in L(v_n)$.
\item \label{item:define-latency} Set $\hat d$ as signal propagation time for the path, where $n_f$ is the refraction index of fiber ($1.62$ \cite{Ramaswami2009}), which is the factor how much slower light travels in a medium , and $c$ the speed of light:

\[ \hat d(x_1,x_2)=\left(\sum_{i=1}^{|l_H|-1} |l_{H,i+1} - l_{H,i}| \right) \cdot c \cdot n_{f}\]
\end{enumerate}

\noindent Some remarks: The definition of $G$ and $H$ depends, among others, on two constants $L_\mathrm{max}$ and $H_\mathrm{max}$. We shall investigate their influence in Section~\ref{sec:as-topology-creation}. Swapping the steps \ref{item:pick-AS-path} and \ref{item:pick-H-path} above (with minor modifications) gives raise to a routing scheme that prioritizes distance over AS hop count. And finally, while we concentrate on latency as induced by signal propagation time and hence geography, modifying step \ref{item:define-latency} would also allow to consider queuing delays (for example) if such information about ASes were available.

\section{AS topology creation}
\label{sec:as-topology-creation}
\subsection{Geographical distribution function}
To generate the initial locations for our model (locations of end devices (function $h$ from
step~\ref{item:pick-candidate-locations} in Section~\ref{sec:adding-latency-graph}) and locations for AS mapping function $L$ (Equation \ref{lgleichung} in Section~\ref{sec:two-graphs})), we need a random distribution that generates the locations. 
And these locations are (very likely) related to population density, which is  not evenly distributed around the world. Hence, we need to check which geographic distribution model for ASes is suitable to pick AS locations.  

We generated an AS location map by querying the contact address publicly available via the WHOIS Services \cite{whoisripe,whoisarin,whoisafrinic,whoisapnic} and the Google Maps Map API \cite{mapsapi}. The plot in Figure~\ref{realasdist} shows the resulting two-dimensional density function.

\begin{figure}
\includegraphics*[width=\columnwidth]{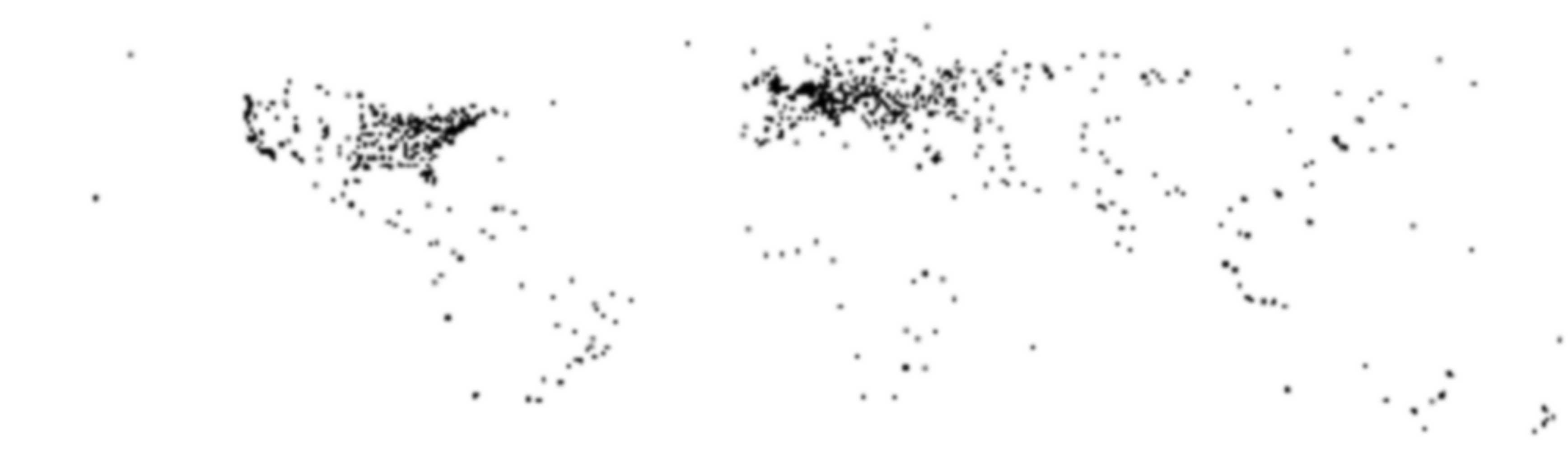}
\caption{Distribution of autonomous systems}
\label{realasdist}
\end{figure}

\subsection{Compactness of ASes}
\label{sec:compactness-ases}
Using a plausible distribution is only one aspect. Another is to reasonably limit the geographic spread of small ASes. An AS with three border routers, one in North America, another in Europa and the third one in Asia is very unrealistic. An AS having three border routers, all of them in one country is, on the other hand,  is quite common. Every AS operator has the goal to minimize costs. The ASes will try to minimize the number links and the costs of links. The longer a network link the higher the cost of this link. 
A cost-efficient AS network will therefore have a smaller average link length between nodes than a cost-inefficient network. Generating a topology that has too many ``long'' links is also unrealistic.

For our network graph generation, we require a formal definition of a valid/realistic AS described in the previous paragraph. The informal formulation suggest that ASes which adhere to a  certain ``compactness'' are valid. 
Using a simple idea like the average distance between nodes will disallow ASes spread over multiple continents since it weighs the links over the ocean too strong. Instead,  we introduce a new compactness metric based on the minimal spanning tree of an AS. Figure~\ref{figmst} illustrates this measure with an example showing a smaller compactness measure for the valid AS.  
Our compactness measure $c(v)$ of an AS $v$ is hence formally defined as the\emph{ average edge length of the minimal spanning tree for the AS} $v$. Let $H_v$ be the complete graph containing all AS locations  $l_i \in L(v)$. Set the edge weight between two locations $l_i$ and $l_j$ as the distance $|l_i - l_j|$. Define $\operatorname{MST}(H_v)$ as the minimal spanning tree of $H_v$. $c(v)$ is then defined as:
\[ c(v) = \frac{\sum_{(l_i,l_j) \in \operatorname{MST} (H_v)} |l_i - l_j|}{n_v -1 }\]

\noindent where $n_v$ $(=|H_v|)$ is the number of geographic locations of this AS $v$. Dividing by $|H_v|$ normalizes this measure to the size of the AS. 
 A cost-efficient network topology will have a smaller spanning tree and therefore a smaller measure $c(v)$ than a costly network topology with unnecessarily  long links. 

\begin{figure}
\includegraphics[width=0.45\columnwidth]{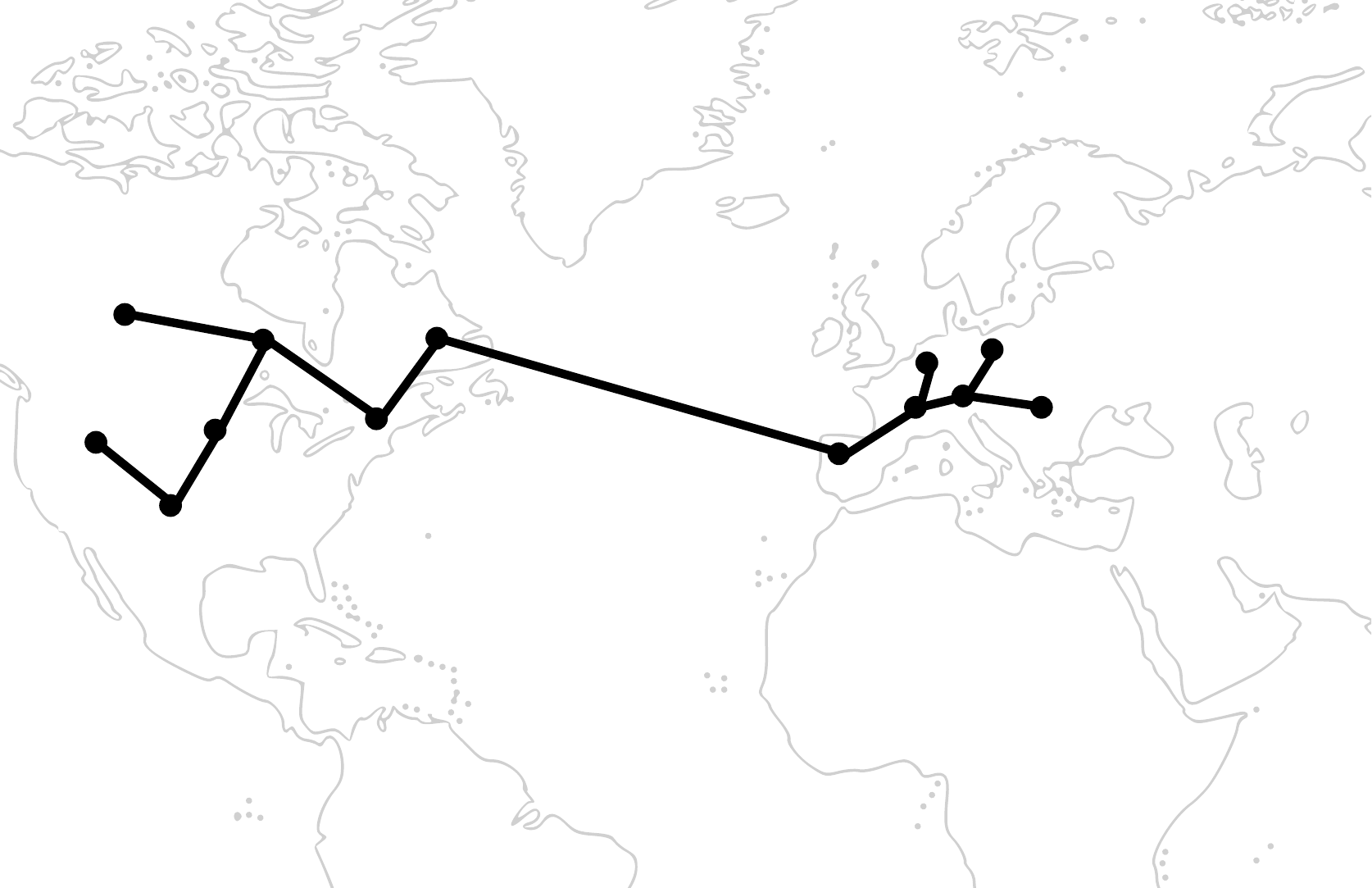}
\hfil
\includegraphics[width=0.45\columnwidth]{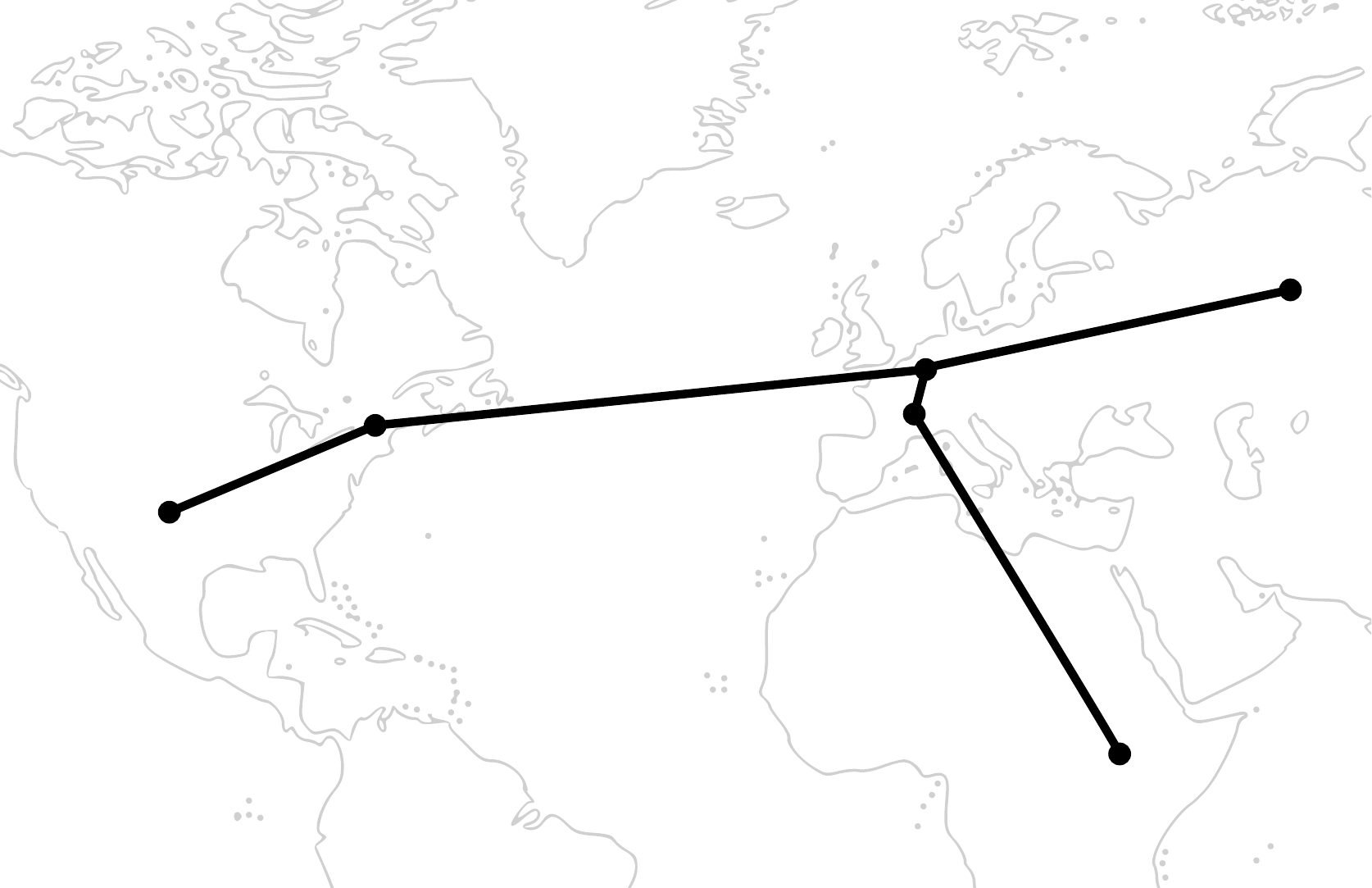}

\caption{AS locations and the induced minimum spanning trees. The left graph has a small average edge length and is a valid placement whereas the right placement is invalid due to its large average edge length. }
\label{figmst}
\end{figure}

Limiting the compactness for all ASes to the same value allows ASes with many locations to be spread further than an AS with a small number of locations. To achieve this, we set an upper bound $c(v) < c_{\max} \ \forall \ v \in G$.

\subsection{Choosing locations}
\label{sec:choos-numb-neighb}

An indicator for the size of an AS $v$ is number of neighbor ASes $\delta(v)$. We set the number of locations higher for large ASes  than for small AS. Let $N$ be the maximal number of locations for an AS and $n$ the minimum number an AS is required to have to gain multiple location. We set the number of locations $n_v$ to:
\[ n_{v} = \max\left(\left\lceil\frac{\delta(u)-n } {\max_{v \in G}(\delta(v))} \cdot N \right\rceil,1\right) \]

After having established the number of locations for each AS
 we generate an initial mapping $L(v)$ with $n_{v}$ random locations, independently drawn according to the two-dimensional  AS density distribution of Figure~\ref{realasdist}. 
Note that the initial $L$ does not guarantee $c(v) < c_{\max}$.  After picking an initial function $L$ we optimise it with following algorithm:
\begin{enumerate}
\item \label{algstep1} Pick uniformly at random two AS locations $l_1$, $l_2$. Set $v_1=L^{-1}(l_1)$, $v_2=L^{-1}(l_2)$. 
\item If $v_1=v_2$ start over
\item Exchange $l_1$ and $l_2$, resulting in a new  $\tilde L=L$. \item Check if $\tilde L$ fulfills $c(v_1) < c_{\max}$ and $c(v_2) < c_{\max}$.
\item Calculate sum of distance to neighbors. Main goal is to eliminate overly long links between neighbors. 
\[ S_{v_i} = \sum_{(v_i,v_j) \in E} \min_{l_i \in l(v_i),l_j \in l(v_j)}\left|l_i - l_j\right|^2  \]

Calculate $S_{v_1}$ and $S_{v_2}$ using $l$, calculate $\tilde S_{v_1}$ and $\tilde S_{v_2}$ using $\tilde L$. If $S_{v_1} + S_{v_2} > \tilde S_{v_1} + \tilde S_{v_2}$ set $L$ to $\tilde L$. 
\item End the algorithm if $L$ has not changed in the last $k$ iterations.
\end{enumerate}
In our experiments the algorithm reached the steady state condition,  in which exchanging locations did not improve $S_{v_1} + S_{v_2}$,  rather  quickly. Although the algorithm does not guarantee the compactness $c(v) < c_{\max}\ \forall v \in G$ in our experiments the condition was always fulfilled (unless setting $c_{\max}$ to very low value). 

\section{Evaluation}
\label{sec:eval-gener-graph}

We are interested to see whether our integrated model can reflect
actual latencies as well as structural network properties, for which
we shall use the number of violations of the triangle inequality. 

\subsection{Empirical latencies}
\label{sec:empirical-latencies}

It would be ideal to compare $\hat{d}$ as it results from our algorithm to the real latency function $d$ of the Internet. Since this function is ultimately unavailable we can compare our results against an approximation of the real latency $d$. To do that, we use latency measurement contained in data sets. For a data set  $S$, define the latency function induced by this data set  $\tilde{d}_S$ as 
\[ \tilde{d}(x_1,x_2) = \begin{cases} \text{ measured latency} & x_1,x_2 \in S \\
\perp & \text{ else} \end{cases} \]

From publicly available data sets measuring the latency between multiple hosts on the Internet, we found the King \cite{IDES} and Meridian \cite{meridian} data sets to be the largest and most used data sets. 
We shall consider how well our modeled latency function $\hat d$ coincides with these measured latencies.

Even though these data sets are the most used data sets (especially by coordinate system researchers) they are far from being flawless. The king data set is built by measuring the latency between DNS servers. We mapped the IP addresses of the DNS servers to geographic locations using a GeoIP service. In Figure~\ref{kingscatterfig}, the geographic distance between a pair of hosts in the data set is plotted against the latency between these hosts.  The dashed line in the graph shows the  theoretical minimum (a fiber optic cable directly connecting both end points).
Even loose verification  shows data points with a latency less than the speed of light. Also, the locations for some of the IP addresses seem to be a little off since for distances smaller than 400 km the latencies are almost random. Reproducing such a behavior with a model allowing only latencies that result from at most the speed of light is not possible. Removing the erroneous data points from the data sets requires to build a model or make assumption about the data.

\begin{figure}
\centering{ 
\includegraphics*[width=0.6\columnwidth] {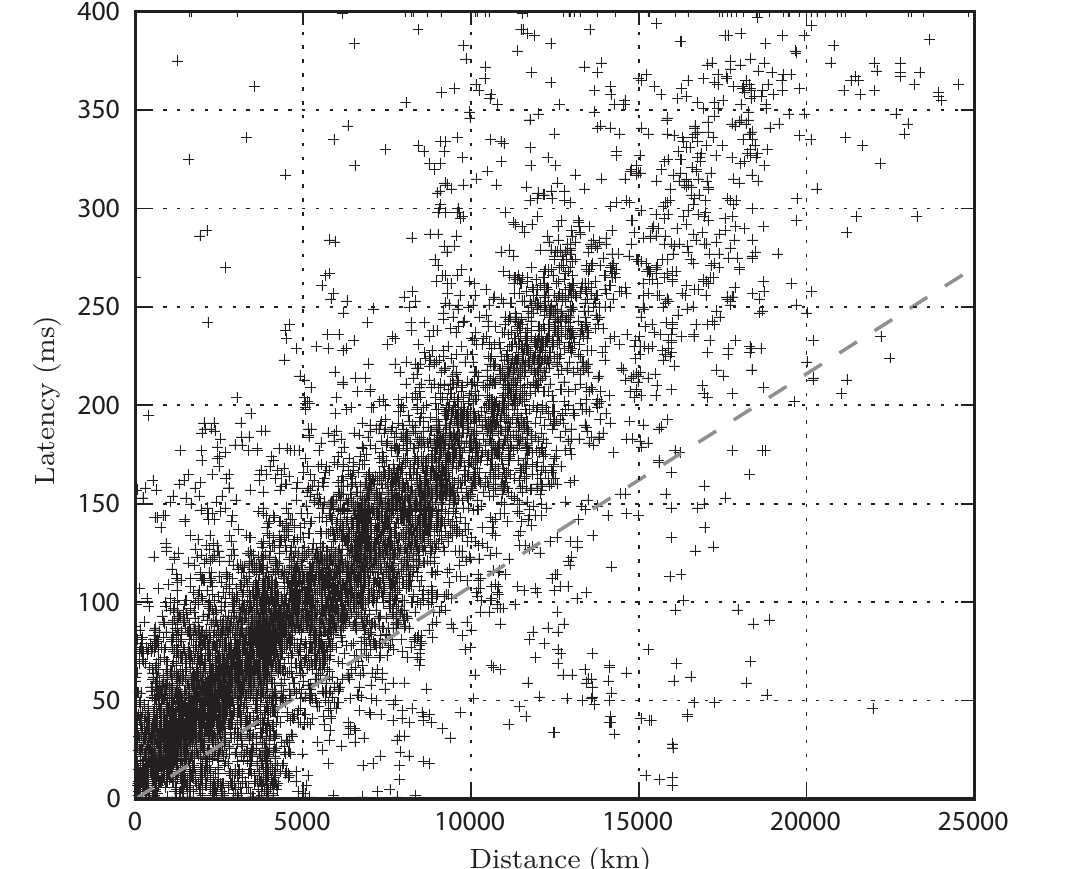}

}
\caption{King data set compared to speed of light in fiber optics (dashed line)} 
\label{kingscatterfig}
\end{figure}

\subsection{Empirical triangle violations}
\label{sec:empir-triangle-viol}

An important metric for network coordinate systems is the triangle violation severity introduced in \cite{Wang2007a}. The idea is give to each end to end connection $(u,v)$ in a set $S$ an indicator for the involvement in triangle violations. A triangle violation is defined as the violation of the metric triangle inequality, which postulates that for a metric $| \cdot |$ the inequality $|x,z| \le |x,y| + |y,z|$ is valid for all $x,y,z$. The TIV severity for host $x_1$ and $x_2 \in X$ in a data set is defined as

\begin{equation}\operatorname{TIV}(x_1,x_2)=\frac{\sum d(x_1,x_2)/(d(x_1,x_i) + d(x_i,x_2))}{|S|}\end{equation}

\noindent for all $x_i \in S$ where $S$ contains all end hosts $x_i$ of the data set which participate in a triangle equality violation with $x_1$ and $x_2$. ($d(x_1,x_2) > d(x_1,x_i) + d(x_i,x_2)$).

The usefulness of the generated topology $H$ and generated latency function $\hat d$ depends on their similarity to the Internet latency. To compare the function $\hat{d}$ with $\tilde{d}$ we will compare the CDF of both function and the TIV. 

\subsection{Parameter choice and results}
\label{sec:param-choice-results}

To generate the top level graph $G$ we decided to use PFP \cite{pfp} for its ability to create an AS graph that matches many properties of the Internet AS graph. We used the optimal parameters identified by the authors of the original paper of  PFP to generate the topology $G$ ($p=0.40,q=0.11$). 

We need to chose the parameters for the model and the algorithm. 
We set the maximum distance between ASes ($L_{\max}$) to 300 km since we do not want AS interconnections to become arbitrarily long and we want to keep the number of interconnections between ASes low. 
We set the limit for $H_{\max}$ to 200 km. This distance corresponds to a maximal added latency of $1$ ms.  The data sets are based on hosts in enterprise or university networks (e.g. DNS server in the King data sets). These hosts typically have a low latency connection to the network. For technology used in consumer technology (xDSL/cable modem) the latency offset will be noticeable but can be modelled by adding a constant to latency inferred from the model.

To find optimal parameters for our embedding algorithm we ran the algorithm with different parameters for $c_{\max}$, $n$ and $N$, which represent the compactness restriction of an AS, the size of an AS to potentially have multiple location and the number of locations the largest AS has. 
As noted in the in Section \ref{sec:choos-numb-neighb} the algorithm does reach a steady state very quickly. We terminated the algorithm after 5000 unsuccessful consecutive iterations. 

After generating model instances we used the locations gained from the geo mapping of the king data to as locations of end devices. Using these end devices we created latency values for our model using the function $\hat{d}$ (the latency function of our model). We then used these latencies to compare our model with the data sets.

We used the  Kolomogorov-Smirnov test (KS test) to compare the model's $\hat d$ with the data sets' $\tilde d$ to select the parameters that give the best approximation of the data set. 
The comparison of the TIV severity using the parameters which have been selected based on for empirical latency CDF. Notice that the graphs diverge at the tails. We believe these tails are artifacts produced by the erroneous data points and other not modeled effects (like queuing delay).

The Figures~\ref{compcdf} and \ref{compcdfmeri} show the comparison of the calculated empirical CDF for the latency of the King and Meridian data set and our generated graphs using the best parameter combination as determined by the aforementioned KS test. As it can be seen, our model very closely approximates the real data set when choosing suitable parameters of our model ($N$, $n$ and $c_{max}$).

\begin{figure}

\includegraphics*[width=0.495\columnwidth]{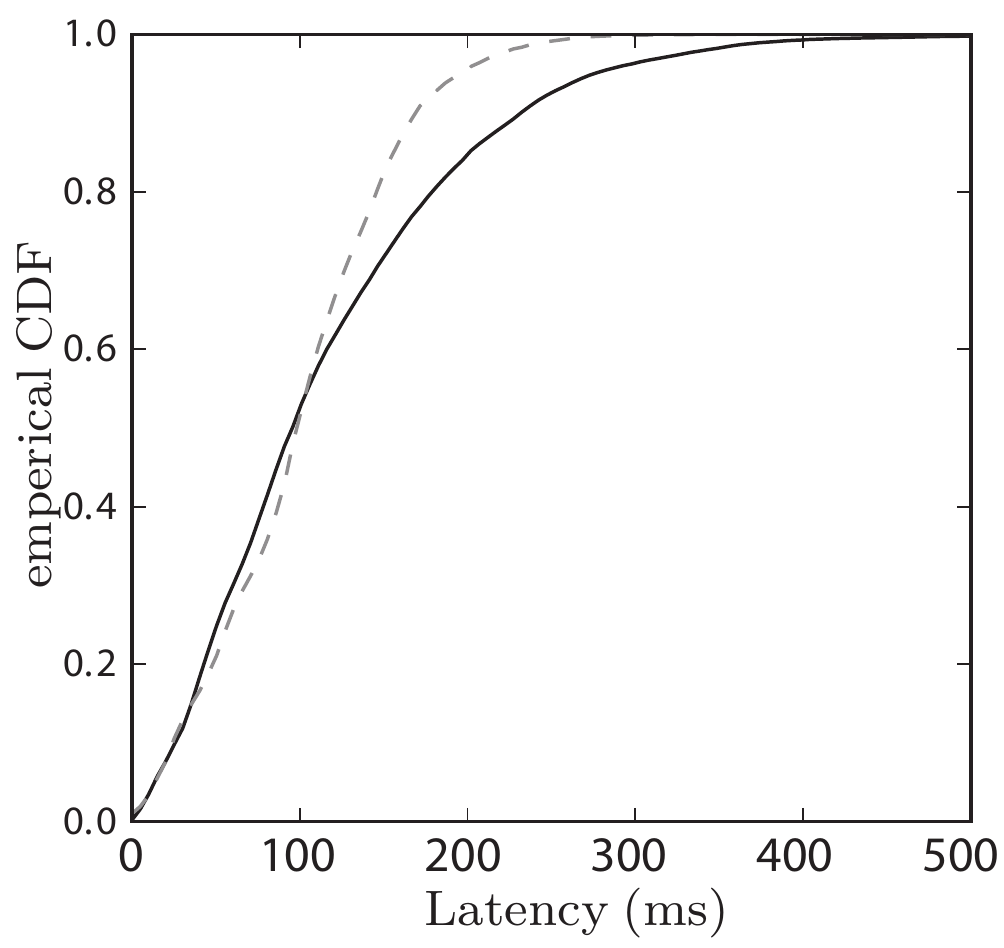}
\includegraphics*[width=0.495\columnwidth]{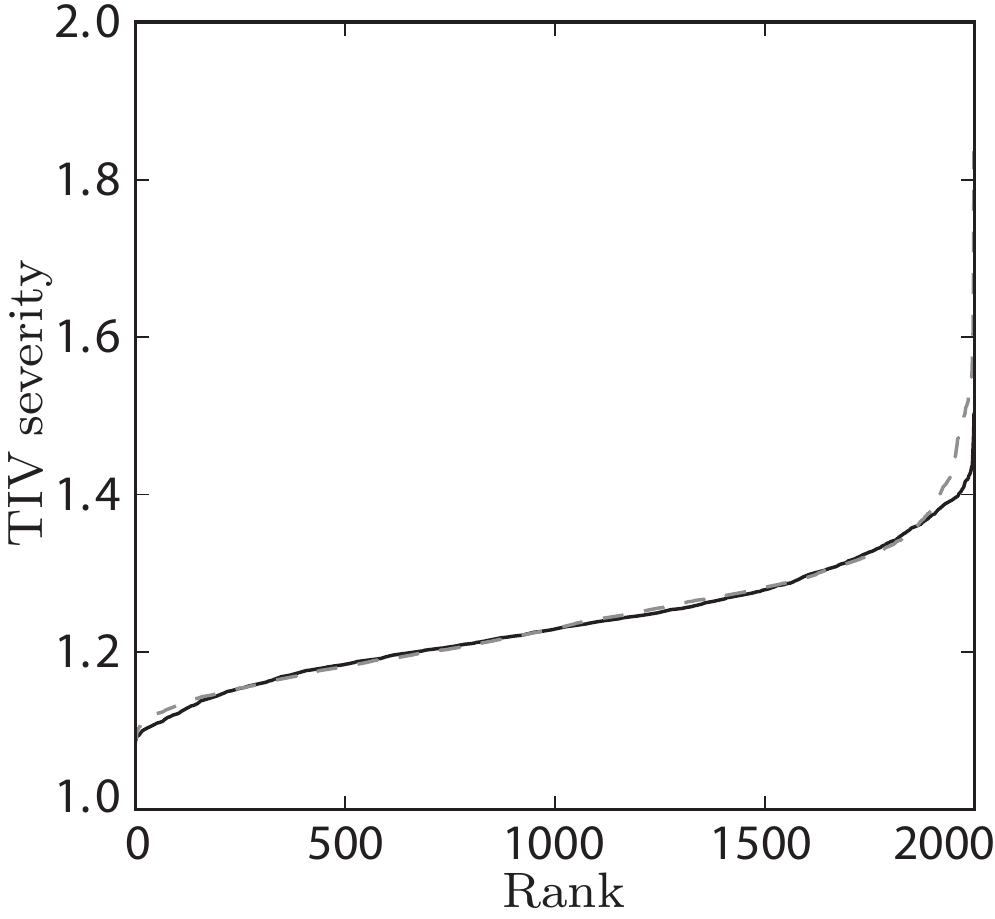}

\caption{Comparison of the the king data set (solid) with our algorithm (dashed, parameters $n=1,N=78.000,c_{\operatorname{max}}=1000)$}
\label{compcdf}
\end{figure}

\begin{figure}

\includegraphics*[width=0.495\columnwidth]{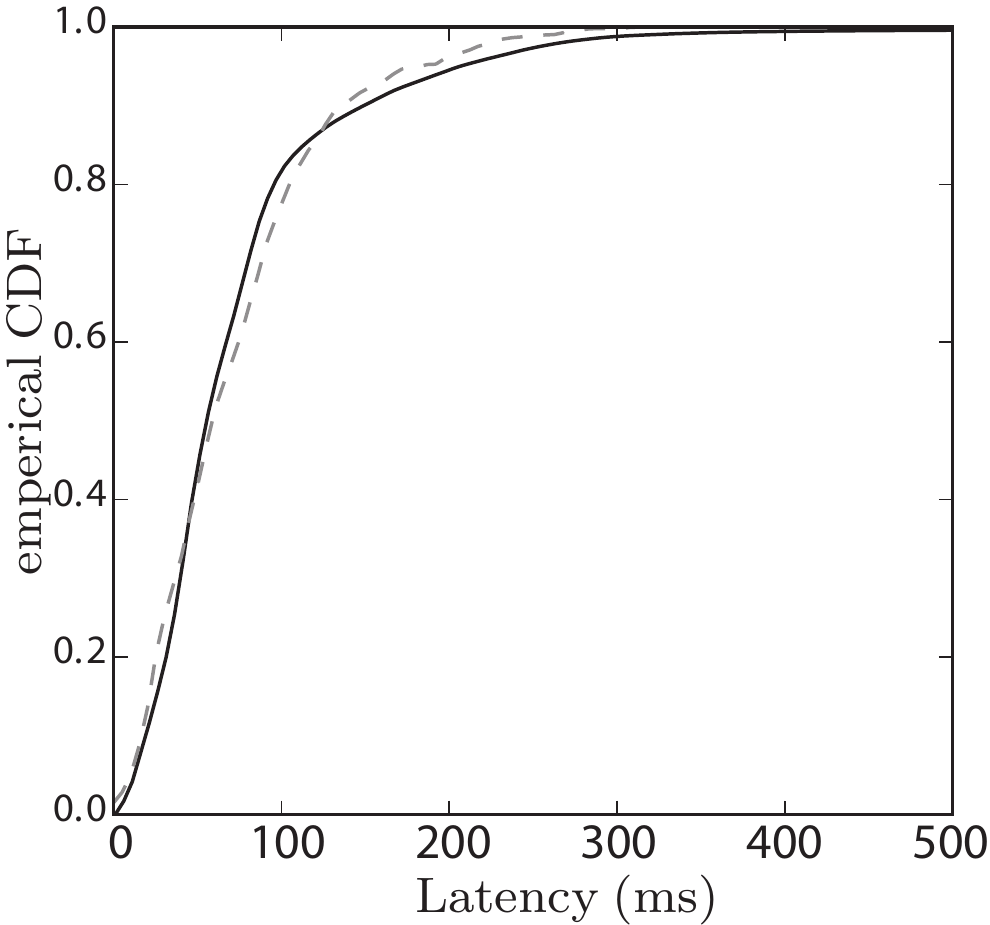}
\includegraphics*[width=0.495\columnwidth]{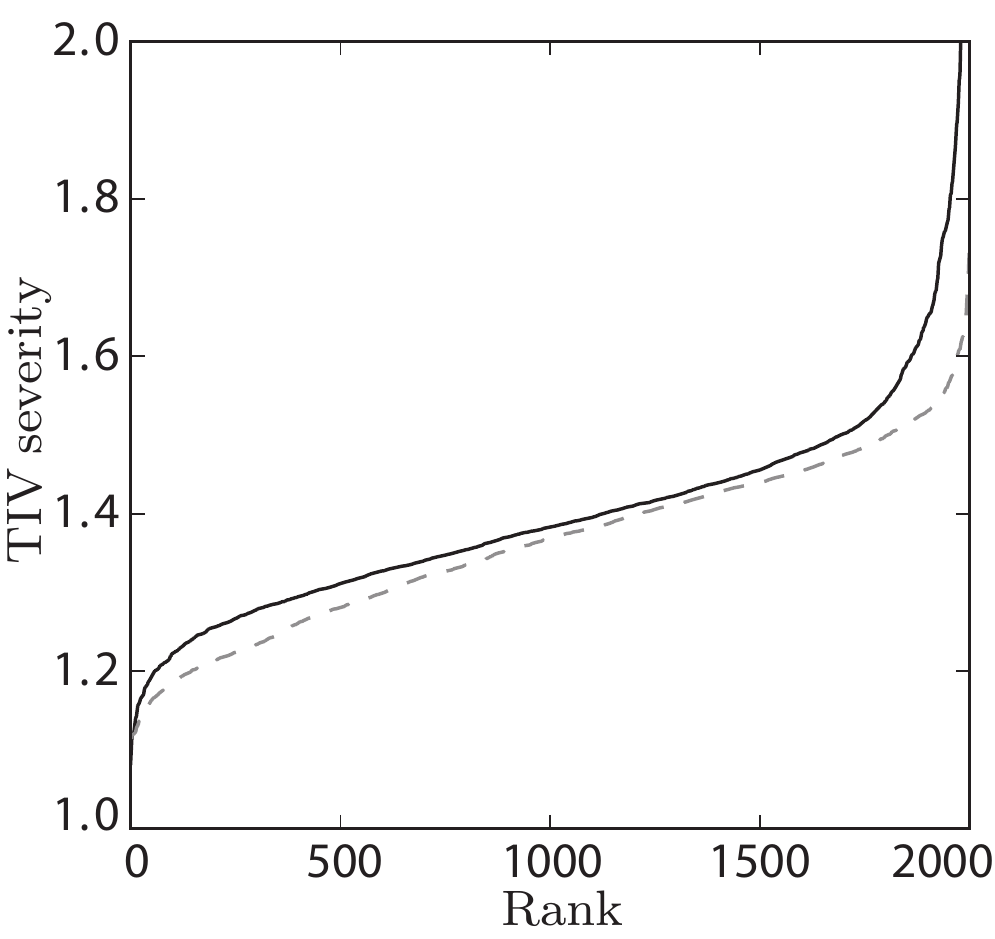}

\caption{Comparison of the meridian data set (solid) with our algorithm (dashed, parameters $n=50,N=36,c_{\operatorname{max}}=2000)$}
\label{compcdfmeri}
\end{figure}

\section{Conclusion}
\label{sec:conclusion}
We have shown that the networks generated by our algorithm reproduce almost the same latency behavior as the reference data sets.

Our approaches recreates a network instead of statistical functions or similar means. Using a network model does not only give a latency but also a network routing path and a network topology. With our approach we can study changes to the routing or the topology and see the resulting on the latency and other network metrics. This makes our approach a useful tool for network research on large when researching the effect of large-scale network applications. Having shown that our can reproduce the characteristics also proves our notion of compactness characterisation of an AS to be a useful asset when creating network topologies.

Our model is unique in the way that it combines simplicity and still gives realistic results. This allows users and researchers of the model to understand and change the model without much effort. 

\section*{Acknowledgment}
This work was partially supported by the German Research Foundation (DFG) within the Collaborative Research Center ``On-The-Fly Computing'' (SFB~901).

\bibliographystyle{abbrv}
\newpage
\bibliography{IEEEabrv,bibliography}

\end{document}